\documentclass[journal]{IEEEtran}
\ifCLASSINFOpdf
  \usepackage[pdftex]{graphicx}
\else
  \usepackage[dvips]{graphicx}
\fi
\begin{document}
\title{Charge and Potential Distributions for Particles Approaching Substrates with Regular Structures}

\author{Wojciech J. Miloch
        and~Sergey V. Vladmirov% <-this % stops a space
\thanks{Manuscript received November 25, 2008.}
\thanks{This work was in part supported by the Norwegian Research Council, NFR, and by the Australian Research Council, ARC.}
\thanks{W. J. Miloch is with the Institute of Theoretical Astrophysics, University of Oslo, Box 1029 Blindern, N-0315 Oslo,
Norway, e-mail: w.j.miloch@astro.uio.no; he is currently visiting the School of Physics, The University of Sydney, Sydney, NSW 2006, Australia.}% <-this % stops a space
\thanks{S. V. Vladimirov is with the School of Physics, The University of Sydney, Sydney, NSW 2006, Australia.}% <-this % stops a space
}
%,~\IEEEmembership{Life~Fellow,~IEEE}


\markboth{IEEE Transactions on plasma science,~Vol.~, No.~, August~2009}%
{Shell \MakeLowercase{\textit{et al.}}: Bare Demo of IEEEtran.cls for Journals}



\maketitle

\begin{abstract}
%\boldmath
The charge and potential distributions for insulating particles approaching a substrate with regular insulating structures are studied by particle-in-cell numerical simulations. An elongated particle and substrate with elongated structures are considered for flowing plasmas. The role of the relative position of the particle and the substrate in their interactions is investigated. It is also demonstrated that the
 interactions are modified by photo\-emission due to directed UV light. The simulations are two dimensional with ions and electrons treated as individual particles. 

\end{abstract}


% \fi
\IEEEpeerreviewmaketitle

\section{Introduction}
\IEEEPARstart{T}{he} plasma aided deposition of particles or grains on substrates allows for the growth of highly ordered structures. Such structures have many applications, including various micro\-electro\-mechanical systems (MEMS) \cite{GadelHak_2006}. MEMS components are typically up to hundred micrometer in size, and the size of the whole device can be of the order of millimeter. In the limiting case, structures of nanometer size can be manufactured, examples of which include nano\-tips, wires, and walls \cite{Ostrikov_Murphy_2007}.

The understanding of interactions between small objects and structures is important for controlling the particle deposition. Many experimental and numerical studies have addressed specific problems \cite{Debleecker_Bogaerts_2004, Bapat_Gatti_2007, Rutkevych_Ostrikov_2007}, but there is still little knowledge on the overall process. In particular, the role of ions and electric fields in the formation of small grain assemblies is not well understood \cite{Ostrikov_2007}, although the corresponding physics is important. In the studies of charging and interactions of dust grains in a plasma it has been shown that the charge, density and potential distributions depend on grain shapes and material, their relative orientations, and the plasma speed \cite{Manweiler_Armstrong_2000, Vladimirov_Maiorov_2003b, Vladimirov_Maiorov_2003a, Miloch_Pecseli_Trulsen_2007}. In the presence of an ion flow, wakes in the ion density and potential as well as a region of enhanced ion density, often referred as the ion focusing, form behind the grain \cite{Maiorov_Vladimirov_2000, Miloch_Pecseli_Trulsen_2008}. In other studies, it has been shown that the grain charge can be effectively modified by UV radiation \cite{Sickafoose_Colwell_2000, Miloch_Vladimirov_2008}. UV light is often used for cleaning purposes prior to the wafer processing or in the photolithography \cite{Ewing_2000}. These processes usually do not include plasma, which can be itself a source of UV radiation.

%nanotechnology and control 
Studies of interactions between objects in flowing plasmas are also important for controlling the dust grain dynamics in plasma processing devices, where dust can contaminate semiconducting wafers \cite{Mankelevich_Olevanov_2008, Selwyn_Heidenreich_1990}. Various dust cleaning and trapping methods are used during processing, but most methods modify also the conditions of the plasma aided particle deposition \cite{Kurimoto_Matsuda_2004}. New ways of the dust control can improve the performance of the process.

Small sizes, different shapes, and large numbers of particles and grains embedded in plasma make it difficult to analyze the problem analytically, while the diagnostics in experiments is limited. Numerical simulations allow for studies of such a system with non-invasive diagnostics.  With an intention to elucidate the physics of interactions between two insulating objects of complex shapes  in flowing plasmas, we present results from numerical simulations on the interactions of an insulating particle (hereafter referred as a grain) approaching an insulating structure with perpendicular rods. We consider  a self-consistent charging of the grain and the substrate in the collision\-less plasma flow. The charging is by collection of electrons and ions. In addition, we allow for photo\-ionization due to directed UV light. The simulated objects are relatively large, i.e., their dimensions are comparable with the electron Debye length $\lambda_{De}$. We note that a similar problem can be formulated for studies of  the coagulation of two insulating dust grains of different sizes as well as for a spacecraft in the low Earth orbit, which is passing through plasma regions of different densities and temperatures \cite{Hastings_1995}.

\section{Numerical code}
The numerical simulations are carried out with the particle-in-cell (PIC) code. The code was previously described in detail in \cite{Miloch_Pecseli_Trulsen_2007, Miloch_Pecseli_Trulsen_2008, Miloch_Vladimirov_2008b}, and only some basic features of the numerical environment are provided here. We consider collision\-less plasma in a  two dimensional system in Cartesian coordinates. Both electrons and ions, which are treated as individual particles, are introduced with Maxwellian velocity distributions within the simulation area of $50\times 50$ in units of the electron Debye length $\lambda_{De}$, with the exception of the areas occupied by the simulated objects. Plasma can leave the simulation area and is also injected through the boundaries at each time step $\Delta t$.  The electron to ion temperature ratio is $T_e/T_i=100$, with $T_e=0.18~\mathrm{eV}$. We have the ion to electron mass ratio $m_i/m_e=120$. The plasma density is $n=10^{10}~\mathrm{m^{-2}}$, and the plasma flow velocity is $v_d=1.25~C_s$, with $C_s$ denoting the speed of sound. Because of the large electron thermal velocity, the plasma flow is represented by the ion drift. The plasma parameters are typical for rf and dc discharge plasma, and are summarized in Table~\ref{tab:input}.

\begin{table}
\caption{Summary of the plasma parameters in the simulation.}
\label{tab:input}
\begin{displaymath}
\begin{array}{ll}
\hline
\hline
 \mathrm{Simulation~area~lengths,}  & L_x=L_y=5 \cdot 10^{-3}\mathrm{~m}\\
\mathrm{Electron~Debye~length,} &  \lambda_{\mathrm{De}}=10^{-4}\mathrm{~m}\\
\mathrm{Electron~temperature,} &  T_e=0.18\mathrm{~eV}\\
\mathrm{Temperature~ratio,} &  T_e/T_i=100\\
\mathrm{Ion~to~electron~mass~ratio,} & m_i/m_e=120\\
\mathrm{Electron~plasma~frequency,} & \omega_{pe}=1.79 \cdot 10^9 \mathrm{~s}^{-1}\\
\mathrm{Ion~plasma~frequency,} & \omega_{pe}=1.63 \cdot 10^8 \mathrm{~s}^{-1}\\
\mathrm{Plasma~density,} & n=10^{10} \mathrm{~m}^{-2}\\
\mathrm{Plasma~drift~speed,} & v_d=1.25~C_s\\
 \hline
\end{array}
\end{displaymath}

\end{table}

\begin{figure}[!t]
\centering
\includegraphics[width=0.6\columnwidth]{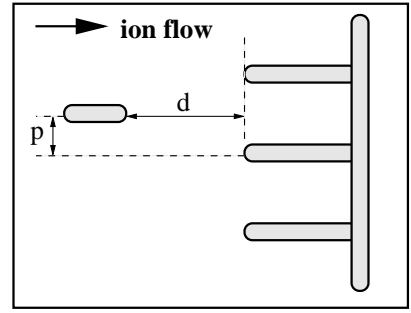}
\caption{Schematics of the numerical arrangement. $d$ is the distance between the grain and a substrate, and $p$ is the offset of the grain from the symmetry line.}
\label{fig:environment}
\end{figure}

The deposited grain and the substrate are placed within the simulation area, far away from the boundaries. The substrate consists of three elongated structures oriented parallel to the flow, and an orthogonal plate. The grain is elongated and aligned with the flow. The grain dimensions are $x \approx 3.6$ and $y \approx 0.75$ measured in units of  $\lambda_{De}$. We consider grains with different distances $d$, and different offsets from the symmetry line $p$ with respect to the substrate: $d=\{1,5,10 \}$, $p = \{0,0.5, 1,2 \}$ both measured in units of $\lambda_{De}$. The schematics of the arrangement is shown in Fig.~\ref{fig:environment}. The deposited grain and substrate are both massive and immobile during the simulation. They are initially not charged. Since we consider insulating objects, the plasma particle hitting the surface of the object remains there for all later times contributing to the local charge density.

When the photo\-emission is considered \cite{Miloch_Vladimirov_2008}, a pulse of a directed light is switched on  at approximately 35 ion plasma periods $\tau_i$ . At this time, we can assume that the surface charge distribution on the objects has reached a stationary level.  The pulse duration is approximately $3 \tau_i$. The code is run typically up to $50\tau_i$. The angle between the incident photons and the plasma flow direction is $\alpha=45^{\circ}$, and the photon flux is $\Psi_{h \nu} = 2.5 \times 10^{19}~\mathrm{m^{-2}s^{-1}}$. When a photon hits the dust, a photo\-electron is produced at distance $l=sv\Delta t$ from the dust surface, where $s$ is an uniform random number $s \in (0,1]$, and $v$ is the photo\-electron speed. Photoelectron velocity vectors are uniformly distributed over an angle of $\pi$ and directed away from the dust surface. We consider mono\-energetic photo\-electrons with the energy $E=0.5~\mathrm{eV}$.

\section{Numerical results}

An insulating grain approaching a substrate is charged by the surrounding plasma, with the total charge on the grain being negative.  We observe positive and negative maxima in the charge and potential distributions on the grain surface. The maxima are due to the ion dynamics around a perfectly insulating object. Behind the grain, the wakes in the plasma potential and density are formed, together with the region of an enhanced ion density, which is due to the bending of ion trajectories by strong electric fields in the vicinity of grain. The charging of an insulating, elongated grain has been studied in more detail elsewhere \cite{Miloch_Vladimirov_2008b}. 

\begin{figure}[!t]
\centering
\includegraphics[width=0.9\columnwidth]{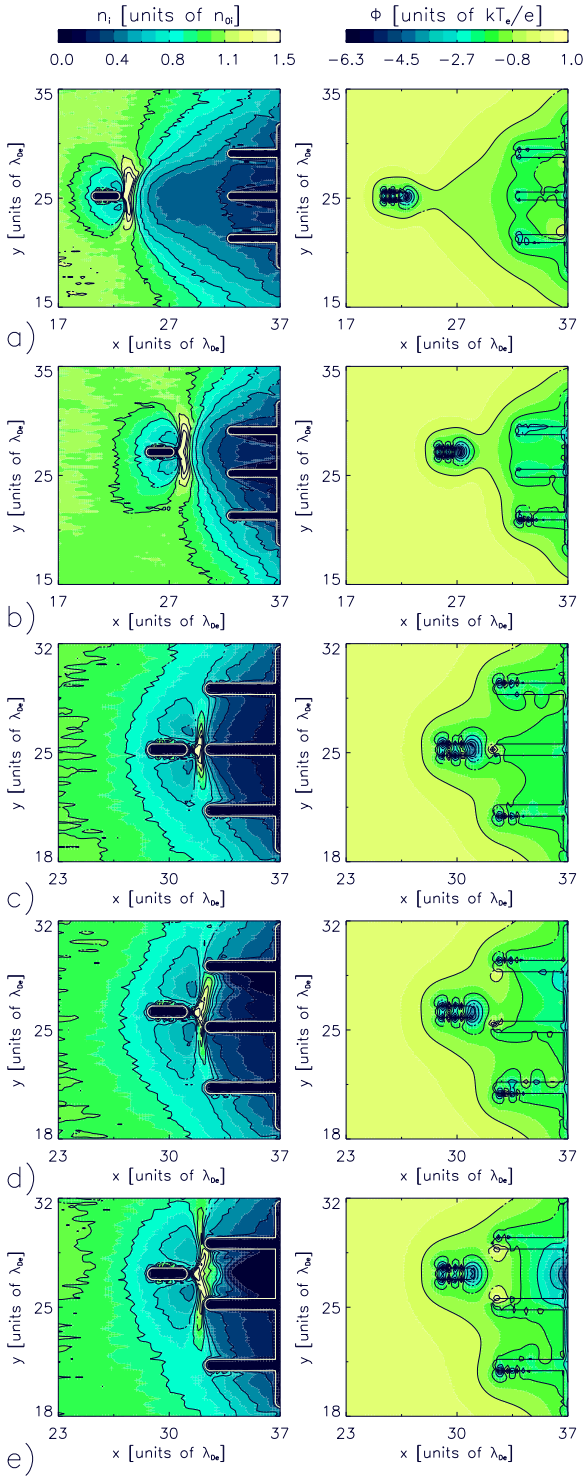}
\caption{Ion density (left) and potential (right) distributions for different distances $d$ and offsets $p$:  (a) $d=10\lambda_{De}$, $p=0$; (b) $d=5\lambda_{De}$, $p=2\lambda_{De}$; (b) $d=1\lambda_{De}$, $p=0\lambda_{De}$, (b) $d=1\lambda_{De}$, $p=1\lambda_{De}$, (b) $d=1\lambda_{De}$, $p=2\lambda_{De}$. The data are averaged over seven ion plasma periods at the end of simulation. Note different scales on $x$ and $y$ axes for (a)-(b) and (c)-(e). Potentials are normalized with $kT_e/e$, where $e>0$. The plasma flows in the positive $x$ direction.}
\label{fig:distance}
\end{figure}

The wake behind the grain modifies the plasma density and potential distributions in the vicinity of a substrate, see Fig.~\ref{fig:distance}. The plasma density is reduced close to the substrate, with the strongest reduction between the rods. With decreasing distance $d$ between the grain and a substrate, the ion focusing region is located closer to the rods on the substrate. For small $d$, the end of the rod placed behind the grain can be exposed to ion fluxes higher than rods located further in the wake of the grain or even in an undisturbed plasma. On the other hand, rods located far from the grain in the direction perpendicular to the flow, can have their ends outside the grain wake for small $d$. In both cases the charges on the rod ends become more positive due to a directed ion flow, and the rod that is closest to the ion focusing region experiences the highest flux. The end of the grain in the shadow of the flow is negatively charged. This can accelerate the grain towards the rods at smaller distances $d$.

An offset $p$ of the grain from the symmetry axis leads to asymmetric charging of the rods on the substrate. The asymmetry is stronger for distances $d=1\lambda_{De}$, when the ion focusing region is close to the rods, see Fig.~\ref{fig:distance} c)-e). For grains located near to one of the rods, most of the ion flux from the ion focus will positively charge the end of this rod. For the grain located between two rods, both rods can receive significant ion flux, which results in both rods having positively charged ends. The charge modification on the rod ends for small $d$ is manifested by variations in the potential distribution on a substrate.  Weaker variations are present also for larger $d$.

\begin{figure}[!t]
\centering
\includegraphics[width=0.8\columnwidth]{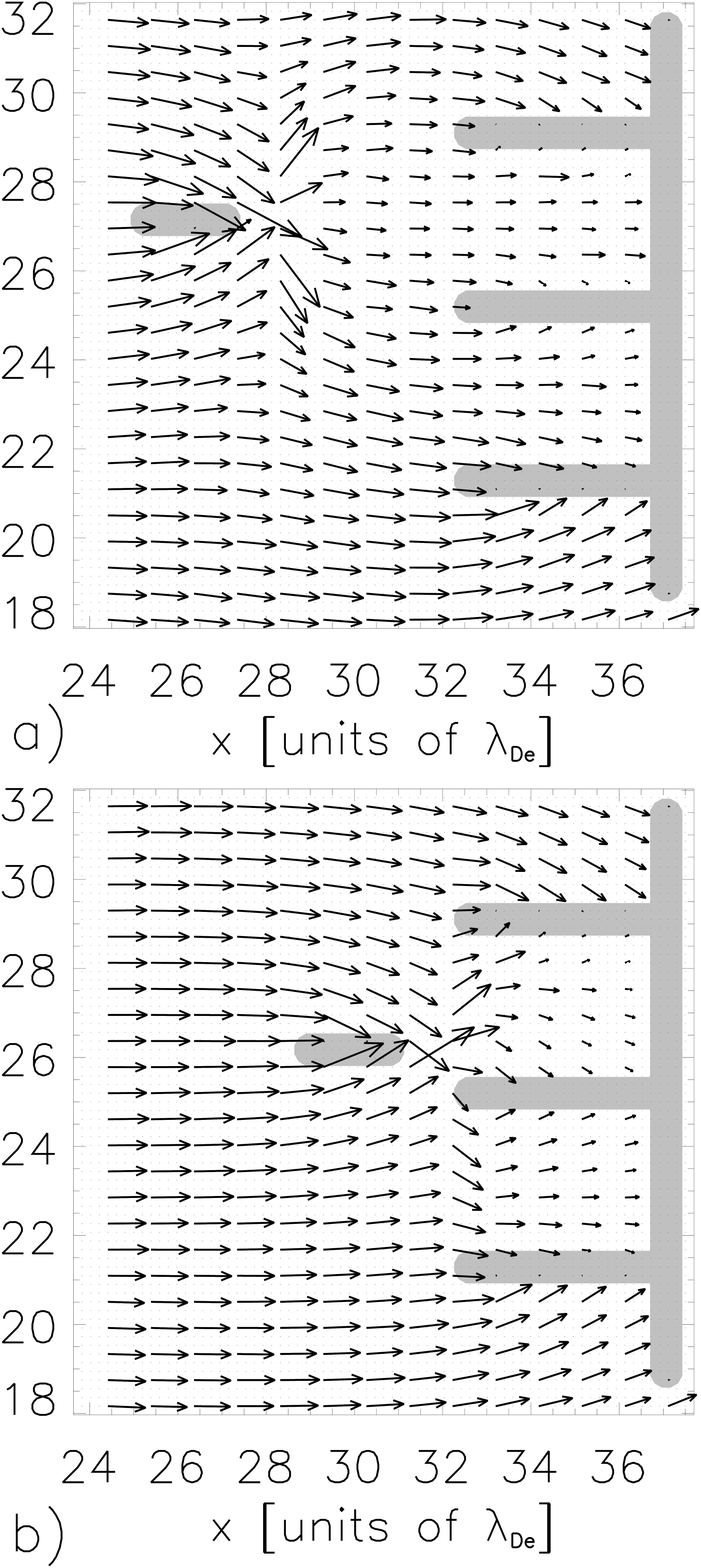}
\caption{The averaged ion flux density for (a) $d=5\lambda_{De}$ and $p=2\lambda_{De}$ and for (b) $d=1\lambda_{De}$ and $p=1\lambda_{De}$. The data are  averaged over a time interval of seven ion plasma periods at the end of simulation and over each grid cell by weighting the ion velocities to the nearest grid points. The ion flux densities are shown for a reduced number of grid points. The areas occupied by the grain and the substrate are marked grey.}
\label{fig:avflux}
\end{figure}

The average ion flux density $\psi_i=\overline{n_{i}}\overline{v_{i}}$  to the rods on the substrate is shown in Fig.~\ref{fig:avflux} for different $d$ and $p$. The average is taken over a time interval of seven ion plasma periods and over the area of each grid cell. We observe an enhancement in the flux for the ion focusing regions. For $d=5\lambda_{De}$ and $p=2\lambda_{De}$, the ion flux density is reduced in the wake and a weak asymmetry is observed. For $d=1\lambda_{De}$ and $p=1\lambda_{De}$ the fluxes to the ends of the two nearest rods are enhanced, but the flux reduction is observed between these rods closer to the substrate. The asymmetry in the flux between the rods closest to the grain is pronounced, while only weak asymmetries are observed between other rods. Some of the ions behind the grain can be lost to the rod surface before reaching the substrate, and thus the wake of a charged insulating grain effectively shields the micro\-channel formed by two rods from a plasma.

\begin{figure}[!t]
\centering
\includegraphics[width=0.8\columnwidth]{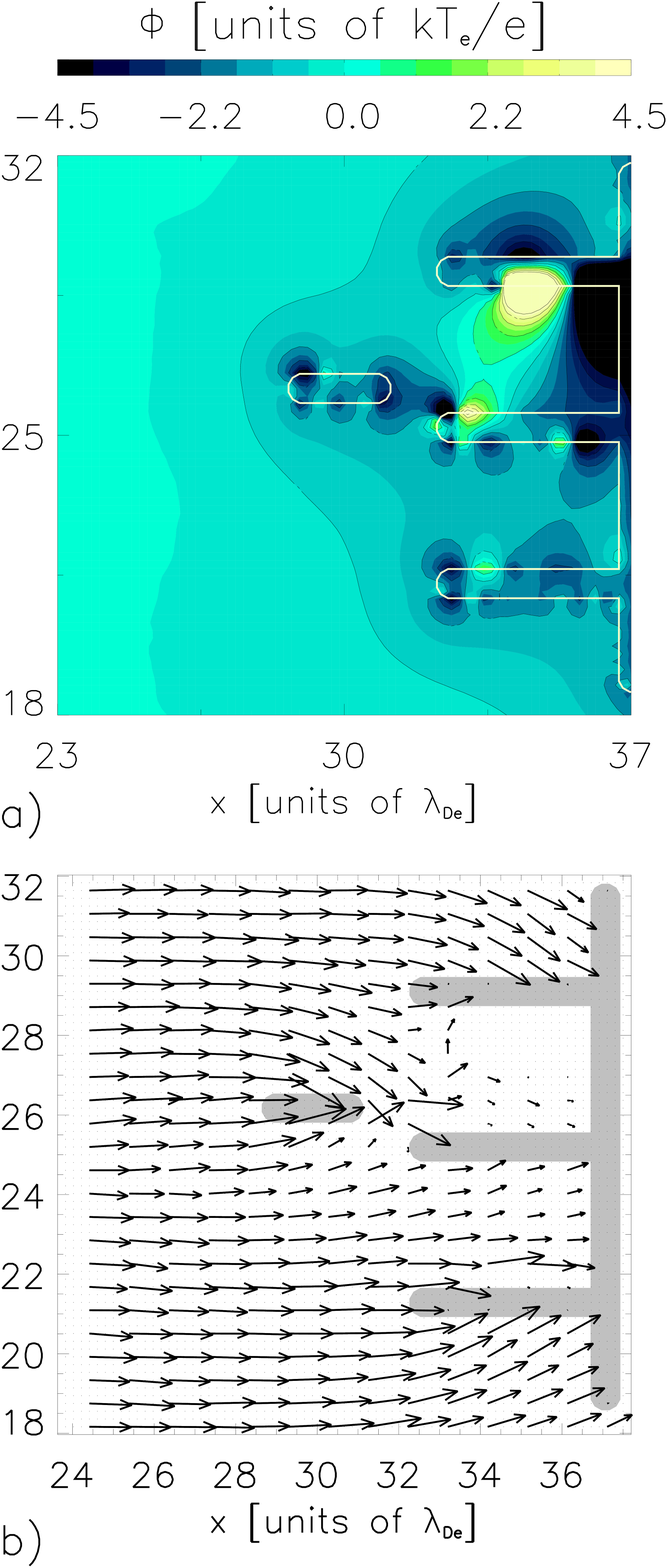}
\caption{The potential distribution (a) and averaged ion flux density (b) averaged over a time interval of $1.5 \tau_i$ at $5.5\tau_i$ after the stop of radiation. $d=p=1\lambda_{De}$.}
\label{fig:photons}
\end{figure}

The surface charge distributions on the grain and a substrate are modified by the photo\-emission due to the pulse of UV light. The grain and substrate become more positively charged during the illumination on the side of the photon incidence. The region of enhanced ion density moves away from the grain and the wake is distorted. The light shadowing by the rods lead to a complicated surface charge distribution. After the pulse, the total charge on each object recovers to the previous value, but the remaining complicated charge distribution due to the photo\-emission can modify the plasma dynamics, see Fig.~\ref{fig:photons}. As a result, the inner sides of the nearest rods can be positively charged: one by photo\-emission, and the other by the ion focusing. The grain between such rods can be drawn towards the substrate. Different angles of the photon incidence can lead to different surface charge distributions and scenarios for the grain deposition.

\section{Discussion and conclusion}
Interactions between an insulating grain and a substrate with insulating structures in a flowing plasma depend on their relative positions. The charge and potential distributions on the substrate are modified by the wake originating from the plasma flowing around the grain. Without photo\-emission, the electrons are Boltzmann distributed, while ions trajectories are bent by strong electric fields in the vicinity of the grain. As a result, the region of an enhanced ion density forms in the wake of the grain. This effect is similar to the electrostatic lensing, and it will also be present for smaller grains \cite{Rutkevych_Ostrikov_2007, Lawson_1988}. The ion focusing is stronger for smaller grains, and weaker focusing is observed also for subsonic ion velocities \cite{Miloch_Pecseli_Trulsen_2008}.

The ion focusing in the wake of the grain has implications for the control of the plasma enhanced chemical vapor deposition (PECVD). The grain above the surface of the substrate electrostatically focuses the ion flux, making it possible to converge the deposited material on a substrate. For elongated grains tilted with respect to the average ion flow, the ion dynamics can be asymmetric behind the grain \cite{Miloch_Vladimirov_2008b}. Therefore, by manipulating the position and orientation of the grain by external forces (e.g., laser and/or electric field), it is possible to control the growth of microstructures on the substrate by changing the PECVD flux locally.

Except for the region of ion focusing, the wake behind the grain is characterized by the reduction in the plasma density. This reduction is observed already at a distance $d=10 \lambda_{De}$. Hence, when a few grains are present close to the substrate, the grain with larger $d$ can influence the dynamics of the grain located closer to the substrate.

The wake structure causes the charge on individual rods on the substrate to be a function of the position of the grain. In particular, the ions streaming out of the ion focusing region contribute to the positive charging of the structures on the substrate. This will lead to attraction of the grain, which is negatively charged on the wake side, by the positively charged ends of the structures. 
Furthermore, complicated surface charge distributions on the grain and substrate will lead to a torque on the grain. The torque will originate not only from the electrostatic potential distribution in the vicinity of the grain, but it is also due to the momentum transfer from incoming ions. By controlling the position of the grain above the substrate, the grain deposition angle can also be controlled. The elongated grain can be, for example, deposited on the structure perpendicular to the flow to further reduce the ion density between the rods.

Photo\-emission due to the pulse of UV light modifies the charge distribution on the grain and substrate and provides yet another mean of controlling the PECVD process. After the pulse, the surface charge distribution remains to be complicated, and it influences the ion dynamics close to the objects. Different scenarios for the grain deposition are possible, depending on the angle of the photon incidence. For instance, the grain can be drawn towards the surface of the substrate between the rods, which can be achieved by the illumination of the substrate from both sides. 

We considered pulsed radiation because the charge saturation is not always the case for insulating grains exposed to continuous radiation. This due to the development of a strong electric dipole moment on such grains. The charge saturates, however, for conducting grains charged by photo\-emission \cite{Miloch_Vladimirov_2008}. For usual conditions  we have the work function for insulators  between five and ten electron\-volts. If the work function of a grain is lower from the substrate, or if the the substrate is biased, the photo\-emission due to directed light allows to modify the grain potential and wake without changing the charge distribution on the substrate for $d > 1 \lambda_{De}$ (when the ion focusing does not modify much the surface charge distribution on the substrate).  In particular, our results for photo\-emission can also be applied for insulating spacecraft components interacting with other objects in space.

Two dimensionality of the model implies that the rods on the substrate can be interpreted as elongated structures forming walls or trenches. The sizes of the simulated objects are comparable with the electron Debye length, thus corresponding to structures of the sizes of MEMS components or larger grains in processing devices. In the limiting case of nanotechnology, most of the processes happen in the sheath region, and the size of the structures is usually much smaller than the Debye length. The physics can be different there due to lack of effective shielding of closely spaced objects. Nevertheless, the characteristic features of the plasma flowing around an object, such as the potential enhancement corresponding to the ion focusing region, will be still present around objects much smaller than the Debye length \cite{Maiorov_Vladimirov_2000, Guio_Miloch_2008}.

To summarize, we have shown that for an insulating grain and insulating substrates, the ion dynamics around the grain will effectively modify the charge on the rods on a substrate and lead to attraction between the rod end and the grain. Photo\-emission changes the charge and potential distributions on the grain, and it can provide a method for controlling the dynamics  of small grains above substrates.

% or


\ifCLASSOPTIONcaptionsoff
  \newpage
\fi



\bibliographystyle{IEEEtran}
%\bibliography{IEEEabrv,bib_ok}


%\begin{thebibliography}{1}


%\end{thebibliography}

%\begin{biography}[{\includegraphics[width=1in,height=1.25in,clip,keepaspectratio]{mshell}}]{Michael Shell}

\begin{IEEEbiography}[{\includegraphics[width=1in,height=1.25in,clip,keepaspectratio]{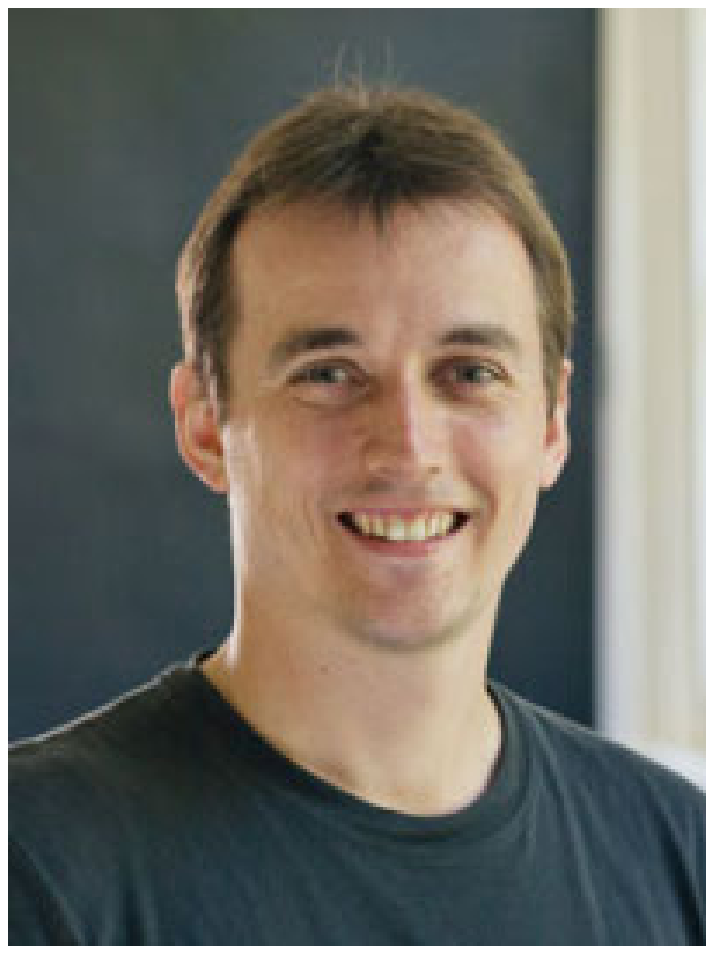}}]{Wojciech J. Miloch}
received his M.Sc. degree in space and plasma physics from the University of Oslo, Norway, in 2006. He is working towards the Ph.D. degree at the Institute of Theoretical Astrophysics, University of Oslo, Norway. He is currently visiting the School of Physics, University of Sydney, Australia.

His research interests include complex plasmas, dust charging, space and astrophysical plasmas, micro\-gravity experiments, and numerical modeling.

\end{IEEEbiography}

\begin{IEEEbiography}[{\includegraphics[width=1in,height=1.25in,clip,keepaspectratio]{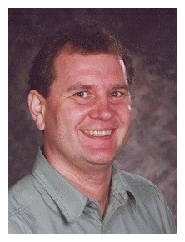}}]{Sergey V. Vladimirov} was born in St. Petersburg, Russia, in 1961. He received 
the Ph.D. degree in theoretical and mathematical 
physics from the P.N. Lebedev Physics Institute, 
Russian Academy of Sciences, Moscow, in 1987 and the Dr. Sci. degree 
in theoretical physics from the General Physics 
Institute, Russian Academy of Sciences, in 1999. 
He is presently an Australian Professorial Fellow 
at the University of Sydney, Australia. His 
main research interests are in nonlinear theory in 
physics, including theoretical plasma physics, plasma turbulence, complex 
plasmas, as well as laboratory plasma applications and space and astrophysical 
plasmas. 

Dr. Vladimirov has received numerous international awards and fellowships 
from A. von Humboldt Foundation and Max Planck Society (Germany), Japan 
Society for the Promotion of Science, Australian Research Council, and Australian Academy of Science. In 2002, he was awarded Pawsey Medal for his 
research in complex plasmas by the Australian Academy of Science.
\end{IEEEbiography}

%\newpage



%\vfill


\end{document}